\documentclass[12pt]{article}\usepackage{quantree2}
\begin{document}

\title{QuanTree and QuanLin,\\
Two Special Purpose\\
Quantum Compilers }

\author{Robert R. Tucci\\
        P.O. Box 226\\
        Bedford,  MA   01730\\
        tucci@ar-tiste.com}

\date{ \today}

\maketitle

\vskip2cm
\section*{Abstract}

This paper introduces
QuanTree v1.1 and QuanLin v1.1,
two Java applications
available for free. (Source code
included in the distribution.)
Each application compiles a different type of
input quantum evolution operator.
The applications
output a quantum circuit
that is approximately equal to the input
evolution operator.
QuanTree
compiles an input evolution operator
whose Hamiltonian is
proportional to the incidence matrix
of a balanced, binary tree graph.
QuanLin compiles an input evolution operator
whose Hamiltonian is
proportional to the incidence matrix
of a line (open string) graph.
Both applications also
output an error, defined as the
distance in the
Frobenius norm between the input evolution
operator and the output quantum circuit.

\section{Introduction}

We say a unitary operator
acting on a set of qubits has been
compiled if
it has been expressed
as a SEO (sequence of elementary
operations, like CNOTs and single-qubit operations).
SEO's are often represented as quantum circuits.

There exist software
(quantum compilers)
like Qubiter\cite{Tuc99}
for compiling arbitrary unitary
operators (operators that have no a priori
known structure).
This paper introduces
 two special purpose quantum
 compilers,
QuanTree and QuanLin.
They are special purpose
in the sense that they can only
compile unitary operators
that have a very definite, special
structure.

QuanTree v1.0 and QuanLin v1.1 are
two Java applications,
available\cite{QuanSuite} for free.
(Source code
included in the distribution.)
Each application compiles a different kind of
input quantum evolution operator.
The applications
output a quantum circuit
that is approximately equal to the input
evolution operator.
QuanTree
compiles an input evolution operator
whose Hamiltonian is
proportional to the incidence matrix
of a balanced, binary tree graph.
QuanLin compiles an input evolution operator
whose Hamiltonian is
proportional to the incidence matrix
of a line (open string) graph.
Both applications also
output an error, defined as the
distance in the
Frobenius norm between the input evolution
operator and the output quantum circuit.

Recently, Farhi-Goldstone-Gutmann (FGG)
wrote a paper\cite{FGG07}
that proposes a quantum algorithm for
evaluating NAND formulas
via a quantum walk over a tree graph
connected to a line (``runway") graph.
Their paper
has inspired a flurry of papers
expanding on their ideas.
Among these papers is one\cite{Theory}
written by me,
which provides all the theoretical
underpinnings of QuanTree and QuanLin.
Please refer to
Ref.\cite{Theory}
and the source code
of QuanTree and QuanLin
if you have any technical
questions that are
no addressed here.
To get a quantum
circuit for the FGG
algorithm requires first
finding a quantum
circuit for the
evolution operators
of a tree and line
graph, which is what
QuanTree and QuanLin do.
A future paper will
combine QuanTree and QuanLin
software to give
a quantum circuit for the
full FGG algorithm.

The standard definition of
the evolution operator
in Quantum Mechanics is
$U= e^{-itH}$, where
$t$ is time and $H$
is a Hamiltonian. Throughout
this paper, we will set
$t = -1$ so $U = e^{iH}$.
If $H$ is proportional
to a coupling constant $g$,
reference to time can be
 restored easily by
replacing the symbol $g$ by
$-tg$, and the symbol $H$ by $-tH$.

\section{QuanTree}

\subsection{Input Evolution Operator}

\begin{figure}[h]
    \begin{center}
    \includegraphics[height=1.5in]{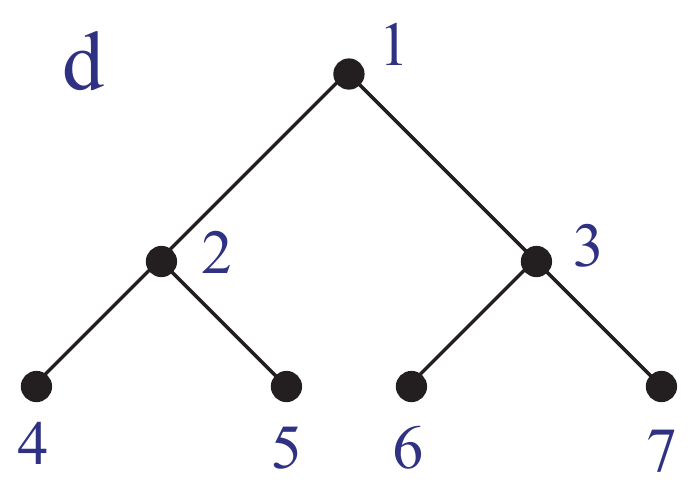}
    \caption{Binary tree with 8 nodes
    }
    \label{fig-tree-graph}
    \end{center}
\end{figure}

A binary tree with $\Lam+1$ levels
has $1+2+2^2 +\ldots 2^{\Lam}=
2^{\Lam+1}-1$ nodes (a.k.a. as states).
To reach $\ns = 2^{\Lam+1}$ states, we
add an extra ``dead"  or ``dud" node,
labelled with the letter ``$d$".
This $d$ node is not connected
to any other node in the graph.
If we include this
dud node, then the number of leaves $\nlvs$
is exactly half the number of nodes:
$\nlvs = \frac{1}{2}2^{\Lam+1}=2^\Lam$.
We will often use $\nb$ for the number of
bits and $\ns=2^\nb$ for the number of states.
Therefore, $\nb = \Lam +1$.
For example, Fig.\ref{fig-tree-graph} shows
a binary-tree
graph with $\Lam=2$ and $\nb=3$. It has
$\ns = 2^{\Lam+1}$=8 nodes,
labelled $d, 1,2, \ldots, 7$,
half of which ($4,5,6,7$) are leaves.

The Hamiltonian
for transitions along the
edges of the graph
Fig.\ref{fig-tree-graph} is:

\beq
H_{tree} = g
\begin{array}{c|c|c|c|c|c|c|c|c|}
&\p{d}&\p{1}&\p{2}&\p{3}&\p{4}&\p{5}&\p{6}&\p{7} \\ \hline
\p{d}& & & & & & & &  \\ \hline
\p{1}& & &1&1& & & &  \\ \hline
\p{2}& &1& & &1&1& &  \\ \hline
\p{3}& &1& & & & &1&1  \\ \hline
\p{4}& & &1& & & & &  \\ \hline
\p{5}& & &1& & & & &  \\ \hline
\p{6}& & & &1& & & &  \\ \hline
\p{7}& & & &1& & & &  \\ \hline
\end{array}
\;,
\label{eq-h-tr}
\eeq
where $g$ is a real number that we will call
the {\bf coupling constant}.
In Eq.(\ref{eq-h-tr}),
empty matrix entries represent zero.

For $\nb = 3$ qubits (i.e.,
$2^\nb = 8$ states),
the {\bf input evolution operator}
for QuanTree
is $U = e^{iH_{tree}}$, where $H_{tree}$
is given by Eq.(\ref{eq-h-tr}).
It is easy to generalize
Fig.\ref{fig-tree-graph} and
Eq.(\ref{eq-h-tr}) to arbitrary $\nb$.
QuanTree can compile
$e^{iH_{tree}}$ for
$\nb\in\{2,3,4,\ldots\}$.

For $r=1,2,3,\ldots$,
if $U = L_r(g) + \calo(g^{r+1})$,
we say
$L_r(g)$ {\bf approximates
(or is an approximant) of order}
$r$
for  $U$.

Given an approximant
$L_r(g) + \calo(g^{r+1})$
of $U$,
and some $\nt= 1, 2, 3\,\ldots$,
one can approximate $U$ by
$\left(L_r(\frac{g}{\nt})\right)^\nt
+ \calo(\frac{g^{r+1}}{\nt^r})$.
We will refer to this as Trotter's trick,
and to $\nt$ as the {\bf number of trots}.

For $\nt=1$, QuanTree
approximates $e^{iH_{tree}}$
with an approximant of
order 3 that is derived
in Ref.\cite{Theory}.
Thus, for $\nt=1$, the error
is $\calo(g^4)$.
For $\nt>1$, the
error is $\calo(\frac{g^4}{\nt^3})$.

\subsection{The Control Panel}

\begin{figure}[h]
    \begin{center}
    \includegraphics[height=4.5in]{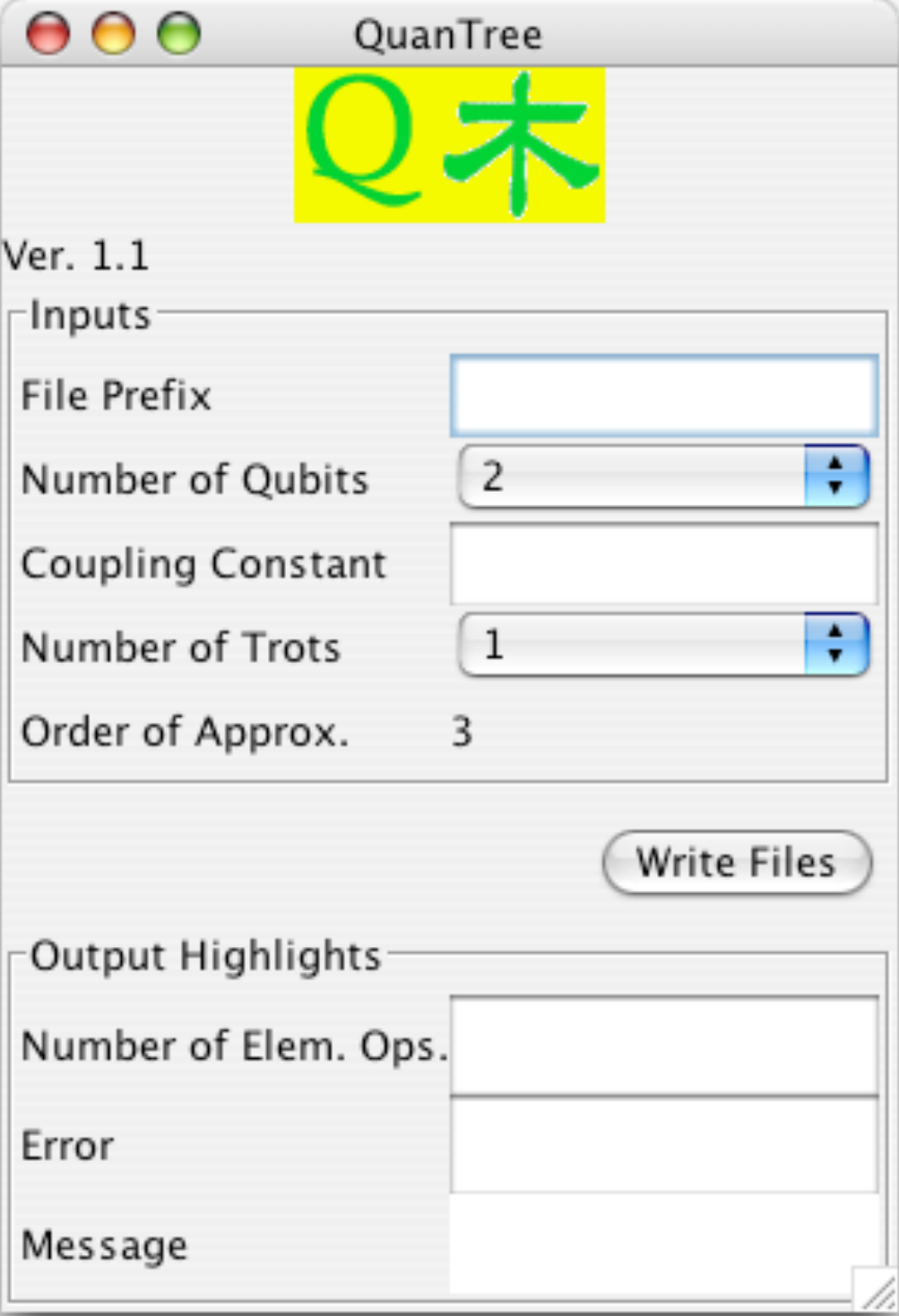}
    \caption{Control Panel of QuanTree}
    \label{fig-qtree-main}
    \end{center}
\end{figure}

Fig.\ref{fig-qtree-main} shows the
{\bf Control Panel} for QuanTree. This is the
main and only window of the application. This
window is
open if and only if the application is running.

The Control Panel
allows you to enter the following inputs:
\begin{description}

\item[File Prefix:] Prefix to the
3 output files
that are written when you press the
{\bf Write Files} button. For example,
if you insert {\tt test} in this text field,
the following 3 files will be written:
\begin{itemize}
\item
{\tt test\_qtree\_log.txt}
(See Fig.\ref{fig-qtree-log})
\item
{\tt test\_qtree\_eng.txt}
(See Fig.\ref{fig-qtree-eng})
\item
{\tt test\_qtree\_pic.txt}
(See Fig.\ref{fig-qtree-pic})
\end{itemize}

\item[Number of Qubits:] The parameter
$\nb=2,3,4,\ldots$ defined above.

\item[Coupling Constant:] The parameter
$g\in \RR$ defined above.

\item[Number of Trots:] The parameter
$\nt=1,2,3,\ldots$ defined above.

\end{description}

The Control Panel displays the
 following
outputs:
\begin{description}

\item[Number of Elementary Operations:]
The number of elementary operations
in the output quantum circuit.
If there are no LOOPs, this is
the number of lines in the English File
(see Sec. \ref{sec-eng-file}), which
equals the number of lines in the
Picture File (see Sec. \ref{sec-pic-file}).
When there are LOOPs, the
``{\tt LOOP k REPS:$\nt$}" and
``{\tt NEXT k}" lines are not counted,
whereas the lines between
``{\tt LOOP k REPS:$\nt$}" and
``{\tt NEXT k}"
are counted $\nt$ times.

\item[Error:] The distance in the
Frobenius norm between the input evolution
operator and the output
quantum circuit (i.e., the SEO given in
the English File).
For a nice review of matrix norms, see
Ref.\cite{Golub}.
For any
matrix $A\in\CC^{n\times n}$, its Frobenius
norm is defined as
$\|A\|_F = \sqrt{\sum_{j,k} A_{j,k}A^*_{j,k}}
$.
Another common matrix norm is the 2-norm.
The 2-norm $\|A\|_2$ of $A$
equals the largest singular value of $A$.
The Frobenius and 2-norm of $A$
are related by\cite{Golub}:
$
\|A\|_2 \leq \|A\|_F \leq \sqrt{2}\|A\|_2
$.
Since the approximant used by QuanTree is
of order 3, if $\epsilon$ denotes the error,
then $\epsilon(g) \approx K g^4$,
for some $K\in \RR$ and $|g|<<1$.
Thus,

\beq
\frac{\log(\epsilon(g_2)/\epsilon(g_1))}
{\log(g_2/g_1)}
\approx 4
\;.
\label{eq-error-order}
\eeq
For example,
with $\nb = 4$ and $\nt = 1$,
QuanTree gives
$\epsilon(0.05) = 1.383\times 10^{-5}$ and
$\epsilon(0.06) = 2.923\times 10^{-5}$, which gives
$\log(\epsilon_2/\epsilon_1)/\log(g_2/g_1)= 4.11$.

\item[Message:]
A message appears in this text field
if you press
{\bf Write Files} with a bad input.
The message tries to explain
the mistake in the input.

\end{description}

\subsection{Output Files}
\begin{figure}[h]
\begin{center}
    \includegraphics[height=3in]{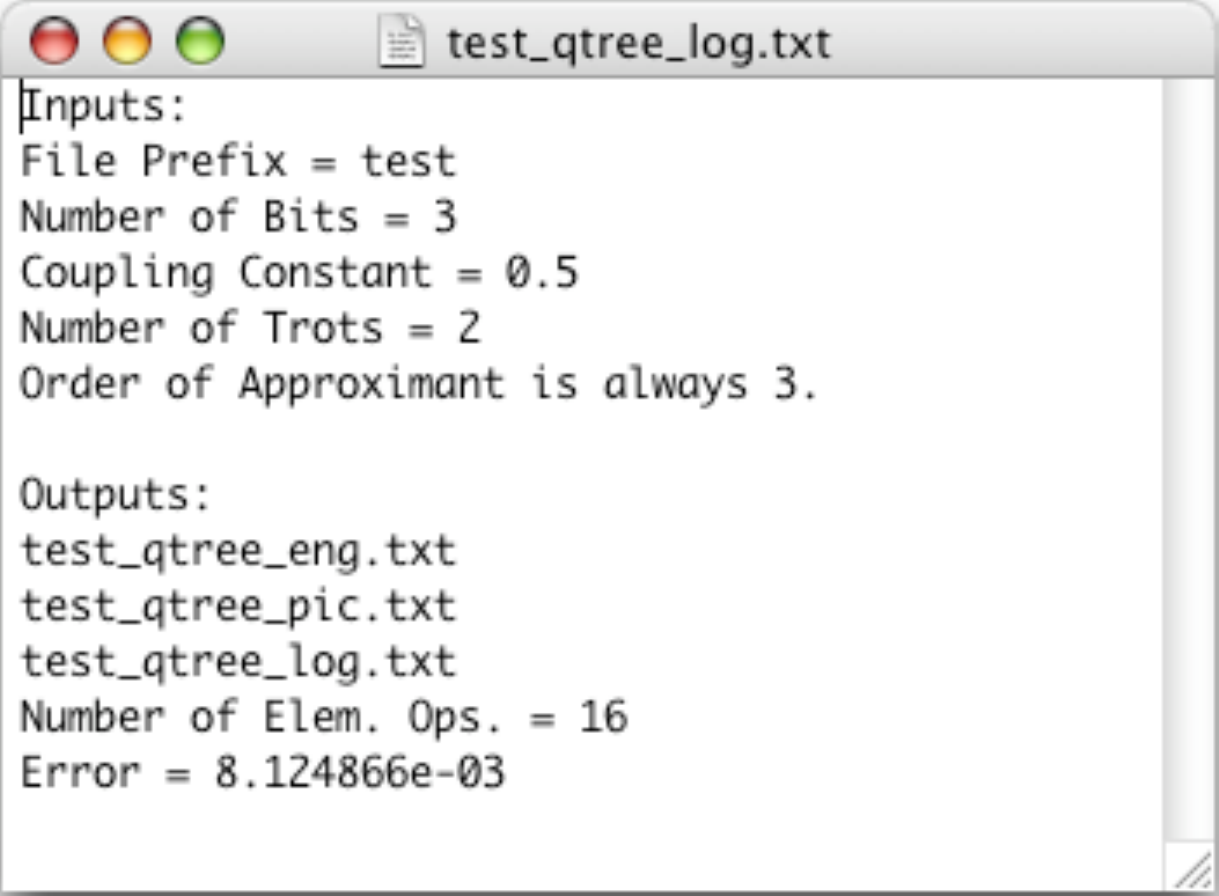}
    \caption{Log File generated by QuanTree}
    \label{fig-qtree-log}
\end{center}
\end{figure}

\begin{figure}[h]
\begin{center}
    \includegraphics[height=3in]{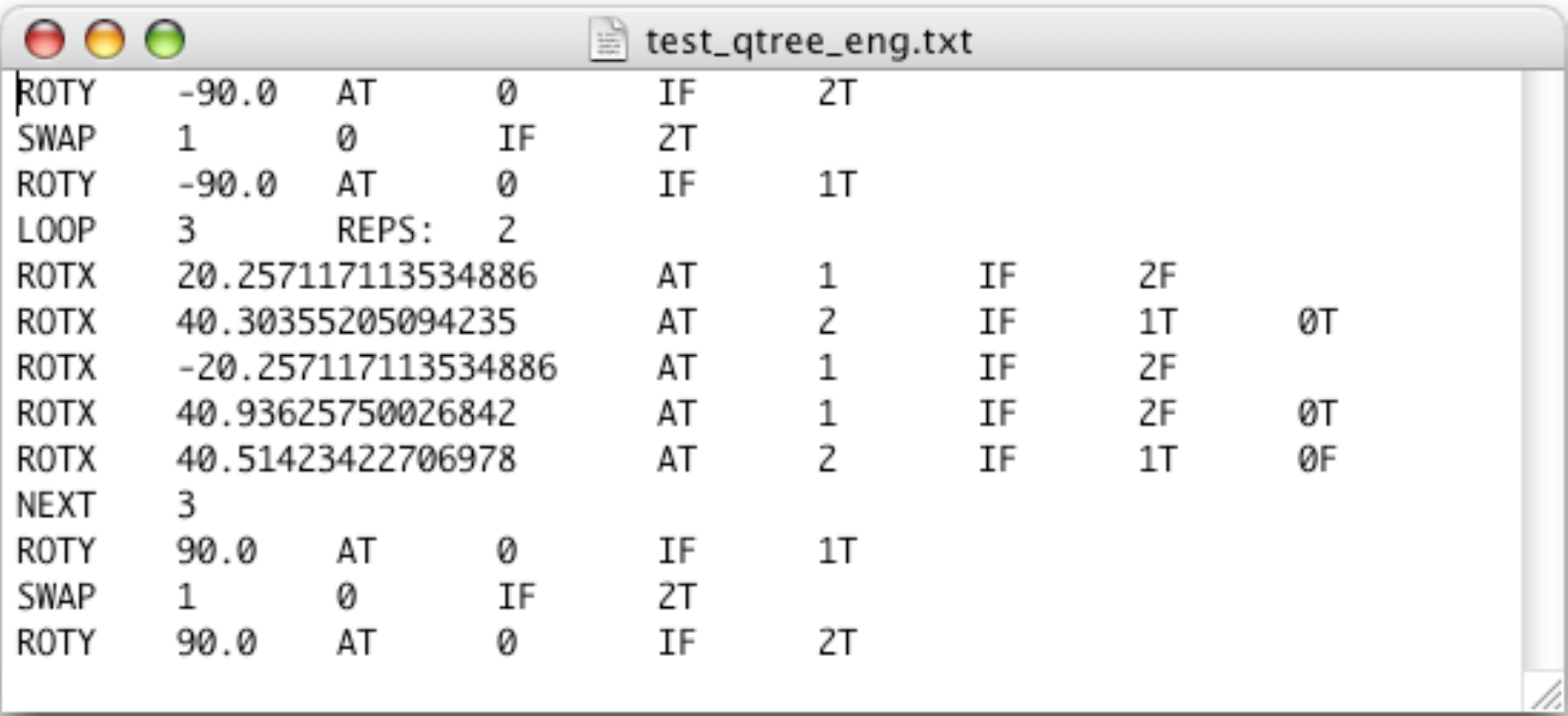}
    \caption{English File generated by QuanTree}
    \label{fig-qtree-eng}
\end{center}
\end{figure}

\begin{figure}[h]
\begin{center}
    \includegraphics[height=3in]{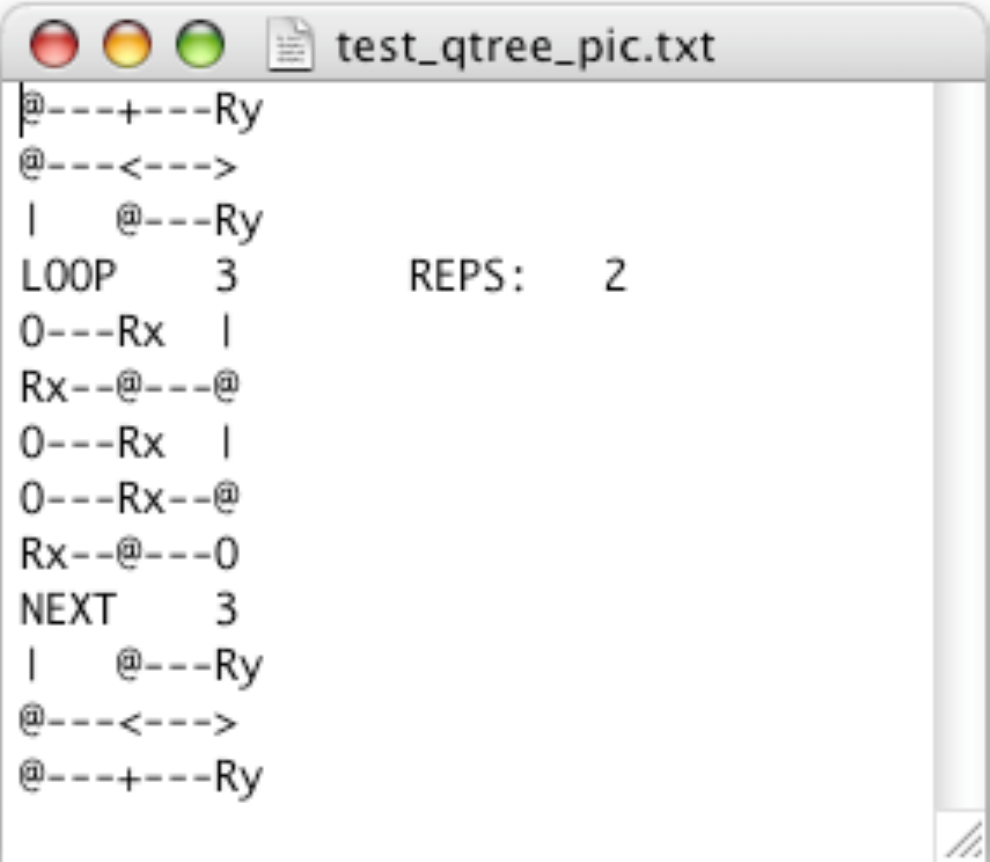}
    \caption{Picture File generated by QuanTree}
    \label{fig-qtree-pic}
\end{center}
\end{figure}

Figs.\ref{fig-qtree-log},
\ref{fig-qtree-eng},
\ref{fig-qtree-pic},
were all generated
in a single run
of QuanTree (by pressing
the {\bf Write Files} button just once).
They are examples of what we call the {\bf
Log File, English File, and Picture File}, respectively, of QuanTree.
Next we explain
the contents of each of these output files.

\subsubsection{Log File}
Fig.\ref{fig-qtree-log}
is an example a Log File.
The Log File
records all the information
found in the Control Panel.

\subsubsection{English File}\label{sec-eng-file}

Fig.\ref{fig-qtree-eng}
is an example of an English File.
The English File
completely specifies the output SEO.
It does so ``in English", thus its name.
Each line represents one elementary operation,
and time increases as we move downwards.

In general, an English File obeys
the following rules:

\begin{itemize}
\item Time grows as we move down the file.

\item Each row
corresponds to one elementary operation.
Each row starts with 4 letters that indicate
the type of elementary operation.

\item For a one-bit operation
acting on a ``target bit" $\alpha$,
the target bit $\alpha$ is given after the word {\tt AT}.

\item If the one-bit operation is controlled, then
the controls are indicated after the word {\tt IF}.
{\tt T} and {\tt F} stand for
true and false, respectively.
{\tt $\alpha$T} stands for
a control $n(\alpha)$ at bit $\alpha$.
{\tt $\alpha$F} stands for
a control $\nbar(\alpha)$ at bit $\alpha$.

\item ``{\tt LOOP k REPS:$\nt$}" and ``{\tt NEXT k}"
mark the beginning and end of $\nt$ Trotter
iterations. {\tt k} labels the loop. {\tt k} also
equals the line-count number (first line is 0)
of the line
``{\tt LOOP k REPS:$\nt$}" in the English file.

\item {\tt SWAP
$\alpha$ $\beta$}
stands for the swap(exchange) operator
$E(\alpha, \beta)$
that swaps
bits $\alpha$ and $\beta$.

\item {\tt PHAS} stands for a controlled
one-bit gate, where the one-bit gate consists of
$diag(1,1)$ times an angle (``phase").

\item {\tt P0PH} stands for a controlled
one-bit gate, where the one-bit gate consists of
$P_0 = \nbar()$ times an angle (``phase").
{\tt P1PH} stands for a controlled
one-bit gate, where the one-bit gate consists of
$P_1 = n()$ times an angle (``phase").

\item {\tt SIGX}, {\tt SIGY},
{\tt SIGZ}, {\tt HAD2}
stand for
the Pauli matrices $\sigx, \sigy, \sigz$
and the one-bit Hadamard matrix $H$.

\item {\tt ROTX}, {\tt ROTY},
{\tt ROTZ}, {\tt ROTN}
stand for
rotations
with rotation axes in the
directions: $x$, $y$, $z$, and
an arbitrary direction $n$, respectively.

\end{itemize}

Here is a list of examples
showing how to translate the mathematical
notation used in Ref.\cite{Theory}
into the English File language:

\begin{tabular}{|l|l|}
\hline
Mathematical language & English File language\\
\hline
\hline
Loop called 5 with 2 repetitions &
{\tt  LOOP 5 REPS: 2}\\
\hline
Next iteration of loop called 5&
{\tt  NEXT 5}\\
\hline
$E(1,0)^{\nbar(3)n(2)}$ &
{\tt SWAP  1  0  IF  3F  2T}\\
\hline
$e^{i 42.7 \nbar(3)n(2)}$ &
{\tt  PHAS 42.7 IF  3F  2T}\\
\hline
$e^{i 42.7 \nbar(3)n(2)}$ &
{\tt  P0PH 42.7 AT  3 IF 2T}\\
\hline
$e^{i 42.7 n(3)n(2)}$ &
{\tt  P1PH 42.7 AT  3 IF 2T}\\
\hline
$\sigx(1)^{\nbar(3)n(2)}$ &
{\tt  SIGX  AT  1  IF  3F  2T}\\
\hline
$\sigy(1)^{\nbar(3)n(2)}$ &
{\tt  SIGY  AT  1  IF  3F  2T}\\
\hline
$\sigz(1)^{\nbar(3)n(2)}$ &
{\tt  SIGZ  AT  1  IF  3F  2T}\\
\hline
$H(1)^{\nbar(3)n(2)}$ &
{\tt  HAD2  AT  1  IF  3F  2T}\\
\hline
$(e^{\frac{i}{2} \frac{\pi}{180} 23.7 \sigx(1)})^{\nbar(3)n(2)}$ &
{\tt  ROTX  23.7  AT  1  IF  3F  2T}\\
\hline
$(e^{\frac{i}{2}  \frac{\pi}{180} 23.7 \sigy(1)})^{\nbar(3)n(2)}$ &
{\tt  ROTY  23.7  AT  1  IF  3F  2T} \\
\hline
$(e^{\frac{i}{2}  \frac{\pi}{180} 23.7 \sigz(1)})^{\nbar(3)n(2)}$ &
{\tt  ROTZ  23.7  AT  1  IF  3F  2T}\\
\hline
$(e^{
\frac{i}{2}  \frac{\pi}{180}
[30\sigx(1)+ 40\sigy(1) + 11 \sigz(1)]
 })^{\nbar(3)n(2)}$ &
{\tt  ROTN  30.0 40.0 11.0  AT  1  IF  3F  2T}\\
\hline
\end{tabular}
\newline\newline

\subsubsection{ASCII Picture File}\label{sec-pic-file}

Fig.\ref{fig-qtree-pic}
is an example of a Picture File.
The Picture File
partially specifies the output SEO.
It gives an ASCII picture of
the quantum circuit.
Each line represents one elementary operation,
and time increases as we move downwards.
There is a one-to-one onto correspondence
between the rows of the English
and Picture Files.

In general, a Picture File obeys
the following rules:

\begin{itemize}
\item Time grows as we move down the file.

\item Each row
corresponds to one elementary operation.
Columns $1, 5, 9, 13, \ldots$ represent
qubits (or, qubit positions). We define the
rightmost qubit  as 0. The qubit
immediately to
the left of the rightmost qubit
is 1, etc.
For a one-bit operator
acting on a ``target bit" $\alpha$,
one places a symbol
of the operator at bit position
$\alpha$.

\item {\tt |} represents a wire
connecting the same qubit at
two times.

\item {\tt -}represents a wire connecting different
qubits at the same time.

\item{\tt +} represents both {\tt |} and {\tt -}.

\item If the one-bit operation is controlled, then
the controls are indicated
as follows.
{\tt @} at bit position $\alpha$ stands for
a control $n(\alpha)$.
{\tt 0} at bit position $\alpha$ stands for
a control $\nbar(\alpha)$.

\item ``{\tt LOOP k REPS:$\nt$}" and ``{\tt NEXT k}"
mark the beginning and end of $\nt$ Trotter
iterations. {\tt k} labels the loop. {\tt k} also
equals the line-count number (first line is 0)
of the line
``{\tt LOOP k REPS:$\nt$}" in the picture file.

\item The swap(exchange) operator
$E(\alpha, \beta)$
is represented by putting
arrow heads {\tt <} and {\tt >} at
bit positions $\alpha$ and $\beta$.

\item A
phase factor $e^{i\theta}$ for some angle
$\theta$ is represented by
placing {\tt Ph} at any bit position
which does not already hold a control.

\item The one-bit gate
$P_0(\alpha) = \nbar(\alpha)$ times an angle
is represented by putting {\tt OP}
at bit position $\alpha$.

\item The one-bit gate
$P_1(\alpha) = n(\alpha)$ times an angle
is represented by putting {\tt @P}
at bit position $\alpha$.

\item One-bit operations
 $\sigx(\alpha)$,
 $\sigy(\alpha)$,
 $\sigz(\alpha)$
and $H(\alpha)$
are represented by placing the letters
{\tt X,Y,Z, H}, respectively,
at bit position $\alpha$.

\item
One-bit rotations
acting on bit $\alpha$,
in the
$x,y,z,n$ directions,
are represented by placing
{\tt Rx,Ry,Rz, R}, respectively,
at bit position $\alpha$.

\end{itemize}

Here is a list of examples
showing how to translate the mathematical
notation used in Ref.\cite{Theory}
into the Picture File language:

\begin{tabular}{|l|l|}
\hline
Mathematical language & Picture File language\\
\hline
\hline
Loop called 5 with 2 repetitions &
{\tt  LOOP 5 REPS:2}\\
\hline
Next iteration of loop called 5&
{\tt  NEXT 5}\\
\hline
$E(1,0)^{\nbar(3)n(2)}$& {\tt 0---@---<--->} \\
\hline
$e^{i 42.7 \nbar(3)n(2)}$ &
{\tt 0---@---+--Ph}\\
\hline
$e^{i 42.7 \nbar(3)n(2)}$ &
{\tt 0P--@\ \ \ |\ \ \ |}\\
\hline
$e^{i 42.7 n(3)n(2)}$ &
{\tt @P--@\ \ \ |\ \ \ |}\\
\hline

 $\sigx(1)^{\nbar(3)n(2)}$& {\tt 0---@---X\ \ \ |} \\
\hline
 $\sigy(1)^{\nbar(3)n(2)}$& {\tt 0---@---Y\ \ \ |} \\
\hline
 $\sigz(1)^{\nbar(3)n(2)}$& {\tt 0---@---Z\ \ \ |} \\
\hline
$H(1)^{\nbar(3)n(2)}$& {\tt 0---@---H\ \ \ |} \\
\hline
$(e^{\frac{i}{2} \frac{\pi}{180} 23.7 \sigx(1)})^{\nbar(3)n(2)}$&
{\tt 0---@---Rx\ \ |} \\
\hline
$(e^{\frac{i}{2}  \frac{\pi}{180} 23.7 \sigy(1)})^{\nbar(3)n(2)}$&
{\tt 0---@---Ry\ \ |} \\
\hline
$(e^{\frac{i}{2}  \frac{\pi}{180} 23.7 \sigz(1)})^{\nbar(3)n(2)}$&
{\tt 0---@---Rz\ \ |} \\
\hline
$(e^{
\frac{i}{2}  \frac{\pi}{180}
[30\sigx(1)+ 40\sigy(1) + 11 \sigz(1)]
})^{\nbar(3)n(2)}$&
{\tt 0---@---R\ \ \ |} \\
\hline
\end{tabular}

\subsection{Behind the Scenes}

Brief summary of
the steps taken by QuanTree
every time you press the {\bf Write Files}
button:
\begin{enumerate}
\item \label{step-eng}
Generate the English
and Picture Files
according to the rules of Ref.\cite{Theory}.
\item Generate $H$.
Calculate the eigenvalues and
eigenvectors of $H$.
Use this eigensystem
 to calculate $U = e^{iH}$.
\item
Read the English File that was
written in Step \ref{step-eng}. Multiply
out the SEO given by the
English File to obtain a unitary
matrix $U'$.
Calculate the error
$\|U-U'\|_F$.
\item Generate the Log File.
\end{enumerate}

\section{QuanLin}
\subsection{Input Evolution Operator}

\begin{figure}[h]
    \begin{center}
    \includegraphics[width=4in]{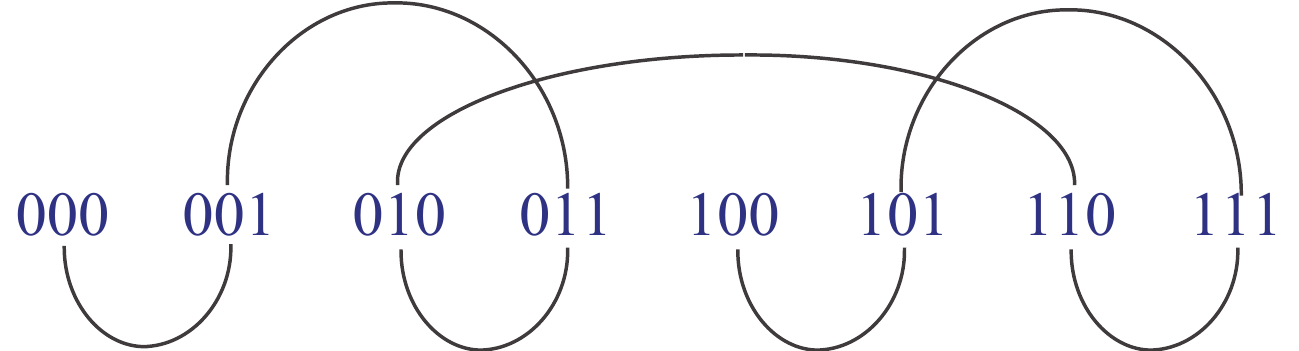}
    \caption{Line (open string) with 8 nodes
    }
    \label{fig-line-graph}
    \end{center}
\end{figure}

Let $\nb$ be the number of bits and
$\ns = 2^\nb$ the number of states.
Fig.\ref{fig-line-graph} shows the
8 possible states for 3 bits. The
states read from left
to right are in increasing ``decimal ordering".
These 8 states can also be ordered
in ``Gray code ordering".
A sequence of Gray code is one wherein
two consecutive states
are labelled by a binary
number and these labels differ only at one bit
position. In Fig.\ref{fig-line-graph},
states connected by an edge (curved line)
are consecutive in a Gray code ordering.

The Hamiltonian
for transitions along the
edges of the graph
Fig.\ref{fig-line-graph} is:

\beq
H_{line} = g
\begin{array}{c|c|c|c|c|c|c|c|c|}
&\p{000}&\p{001}&\p{010}&\p{011}&\p{100}&\p{101}&\p{110}&\p{111} \\ \hline
\p{000}&0&1& & & & & &  \\ \hline
\p{001}&1&0& &1& & & &  \\ \hline
\p{010}& & &0&1& & &1&  \\ \hline
\p{011}& &1&1&0& & & &  \\ \hline
\p{100}& & & & &0&1& &  \\ \hline
\p{101}& & & & &1&0& &1  \\ \hline
\p{110}& & &1& & & &0&1  \\ \hline
\p{111}& & & & & &1&1&0 \\ \hline
\end{array}
\;,
\label{eq-h-line}
\eeq
where $g$ is a real number that we will call
the {\bf coupling constant}.
In Eq.(\ref{eq-h-line}),
empty matrix entries represent zero.

For $\nb = 3$ qubits (i.e.,
$2^\nb = 8$ states),
the {\bf input evolution operator}
for QuanLin
is $U = e^{iH_{line}}$, where $H_{line}$
is given by Eq.(\ref{eq-h-line}).
It is easy to generalize
Fig.\ref{fig-line-graph} and
Eq.(\ref{eq-h-line}) to arbitrary $\nb$.
QuanLin can compile
$e^{iH_{line}}$ for
$\nb\in\{2,3,4,\ldots\}$.

For $r=1,2,3,\ldots$,
if $U = L_r(g) + \calo(g^{r+1})$,
we say
$L_r(g)$ {\bf approximates
(or is an approximant) of order}
$r$
for  $U$.

Given an approximant
$L_r(g) + \calo(g^{r+1})$
of $U$,
and some $\nt= 1, 2, 3\,\ldots$,
one can approximate $U$ by
$\left(L_r(\frac{g}{\nt})\right)^\nt
+ \calo(\frac{g^{r+1}}{\nt^r})$.
We will refer to this as Trotter's trick,
and to $\nt$ as the {\bf number of trots}.

For $\nt=1$, QuanLin
approximates $e^{iH_{line}}$
with a Suzuki approximant of
order $r=2, 4, 6, \ldots$
that is derived
in Ref.\cite{Theory}.
Thus, for $\nt=1$, the error
is $\calo(g^{r+1})$.
For $\nt>1$, the
error is $\calo(\frac{g^{r+1}}{\nt^r})$.

\subsection{The Control Panel}

\begin{figure}[h]
\begin{center}
    \includegraphics[height=4.5in]{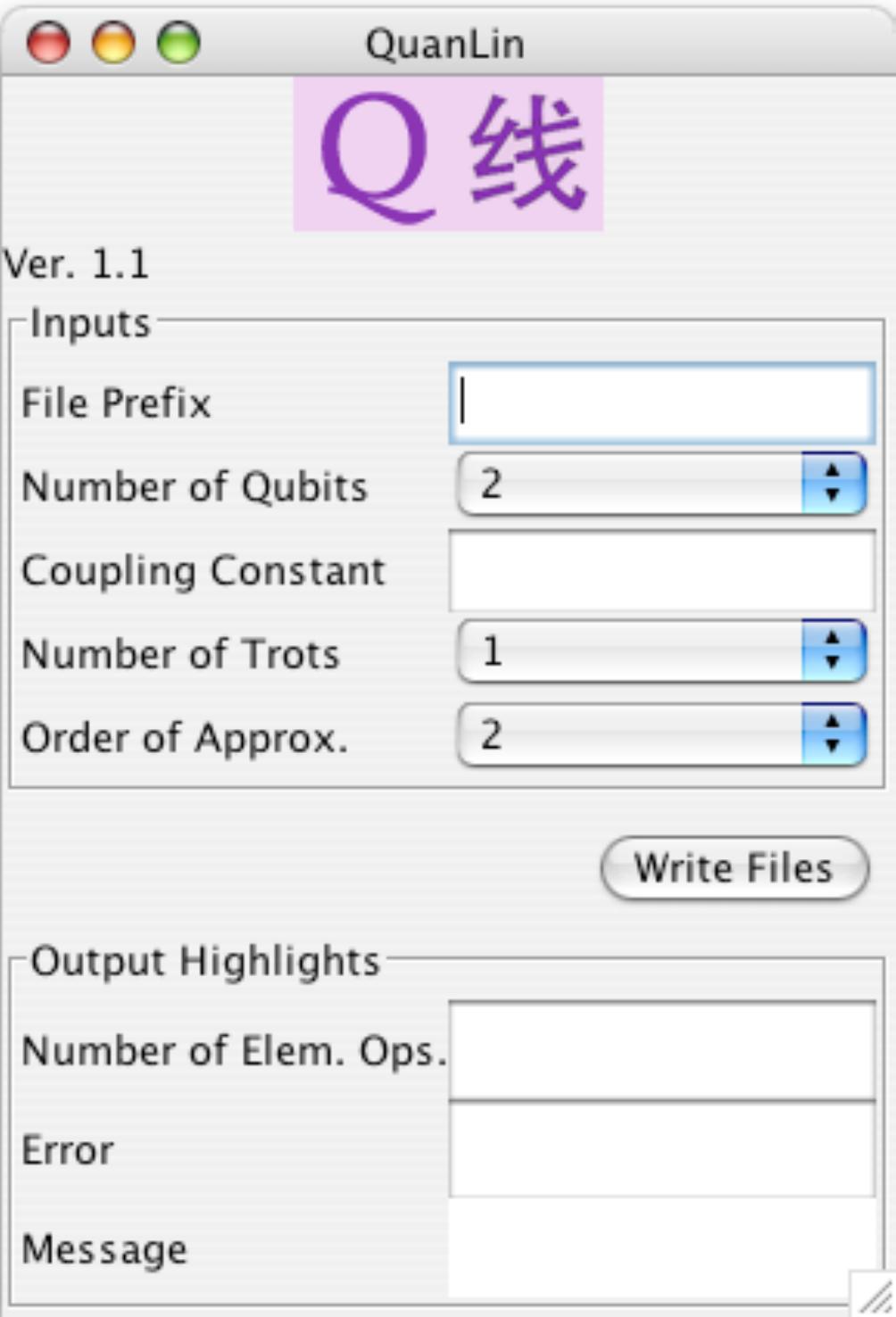}
    \caption{Control Panel of QuanLin}
    \label{fig-qline-main}
\end{center}
\end{figure}

Fig.\ref{fig-qline-main} shows the
{\bf Control Panel} for QuanLin. This is the
main and only window of the application. This
window is
open if and only if the application is running.

The Control Panel
for QuanLin is almost identical
to that for QuanTree.
The significance of the various
data fields in the
Control Panel for
QuanLin is the same as for
QuanTree.

\subsection{Output Files}

\begin{figure}[h]
\begin{center}
    \includegraphics[height=3in]{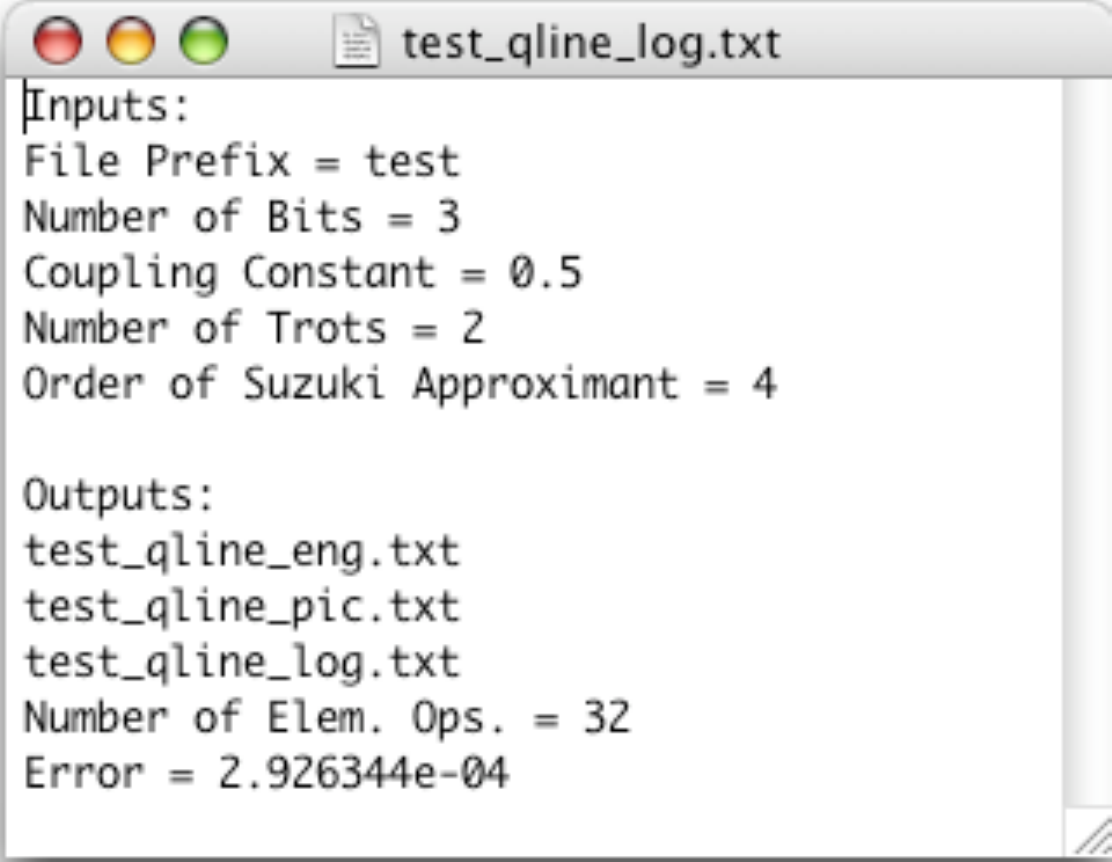}
    \caption{Log File generated by QuanLin}
    \label{fig-qline-log}
\end{center}
\end{figure}

\begin{figure}[h]
\begin{center}
    \includegraphics[height=4in]{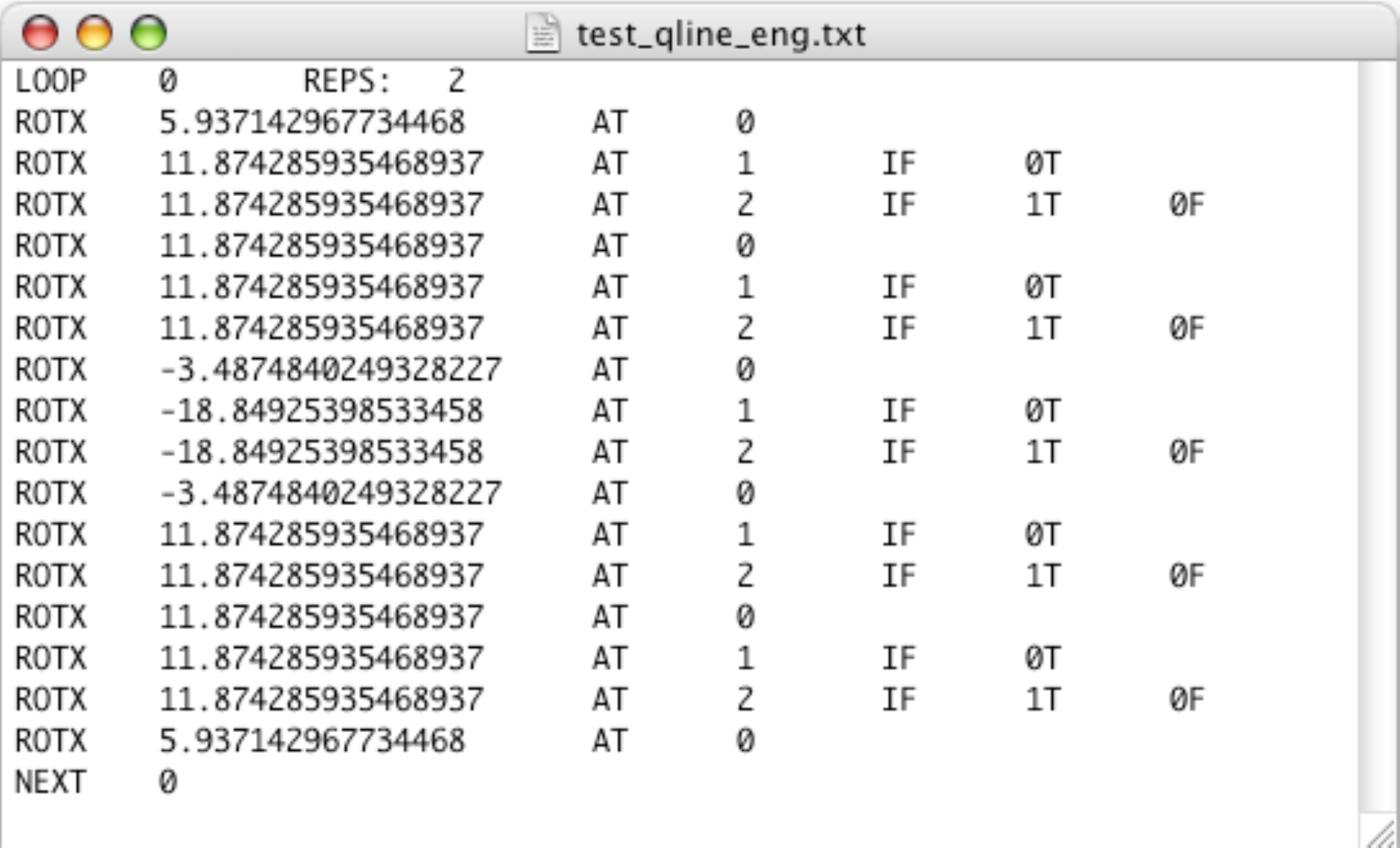}
    \caption{English File generated by QuanLin}
    \label{fig-qline-eng}
\end{center}
\end{figure}

\begin{figure}[h]
\begin{center}
    \includegraphics[height=4in]{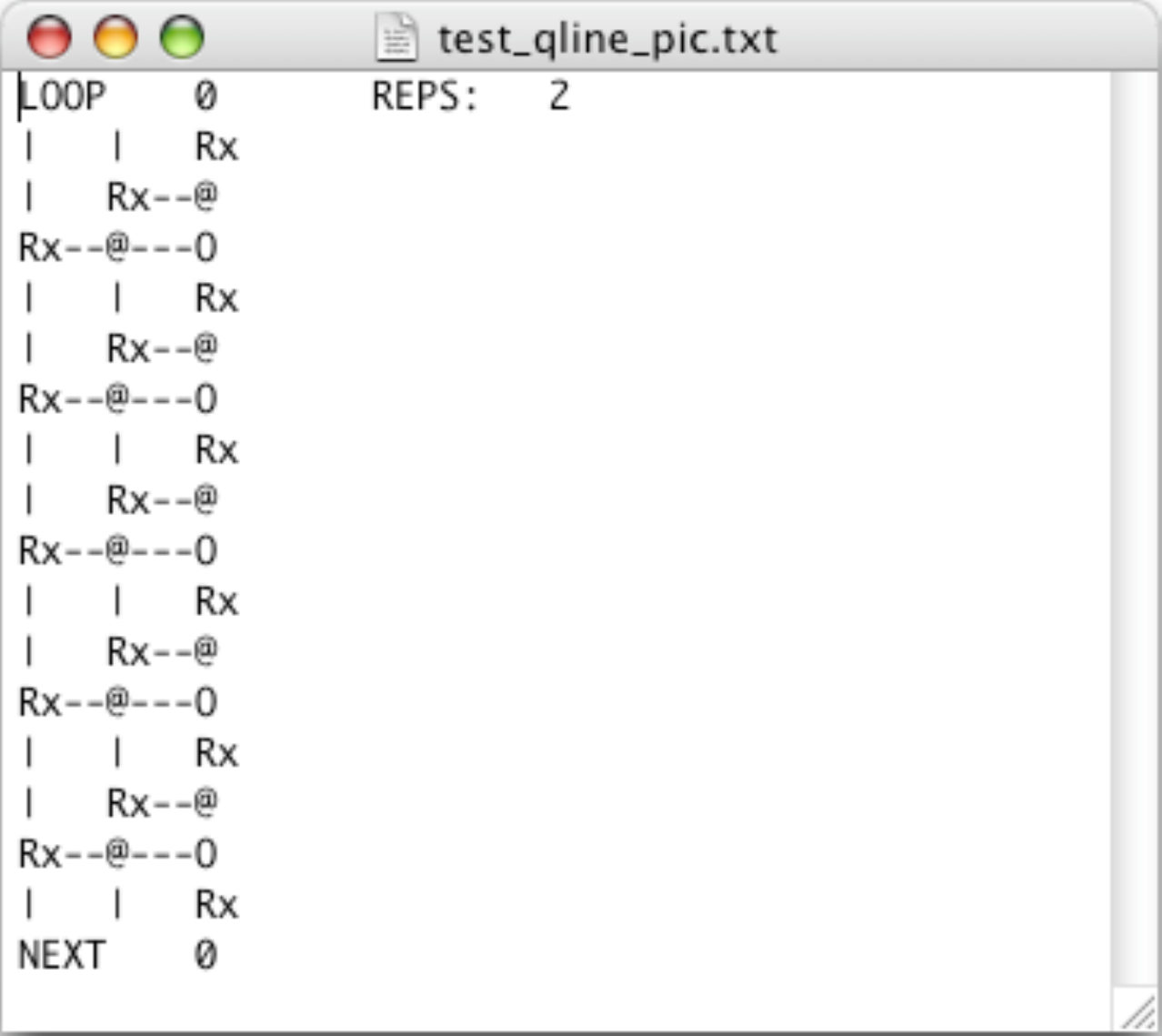}
    \caption{Picture File generated by QuanLin}
    \label{fig-qline-pic}
\end{center}
\end{figure}

Figs.\ref{fig-qline-log},
\ref{fig-qline-eng},
\ref{fig-qline-pic},
were all generated
in a single run
of QuanLin (by pressing
the {\bf Write Files} button just once).
They are examples of what we call the {\bf
Log File, English File, and Picture File},
respectively, of QuanLin.
These files are analogous
to their namesakes for QuanTree.
They follow the same rules.

\end{document}